\documentclass[twocolumn]{aastex7}
\usepackage{hyperref}

\journalinfo{The Astrophysical Journal Letters (accepted)}

\begin{document}

\title{Zeeman Doppler Imaging of $\tau$\,Ceti: The Weakest Magnetic Field Detected in a Sun-like Star}

\author[0000-0001-5180-2271]{Federica Chiti}
\affiliation{Institute for Astronomy,
University of Hawai‘i at Mānoa,
2680 Woodlawn Dr., Honolulu, HI 96822, USA}
\email{fchiti@hawaii.edu}

\author[0000-0003-3061-4591]{Oleg Kochukhov}
\affiliation{Department of Physics and Astronomy, Uppsala University, Box 516, SE-75120 Uppsala, Sweden}
\email{oleg.kochukhov@physics.uu.se}

\author[0000-0002-4284-8638]{Jennifer L. van Saders}
\affiliation{Institute for Astronomy,
University of Hawai‘i at Mānoa,
2680 Woodlawn Dr., Honolulu, HI 96822, USA}
\email{jlvs@hawaii.edu}

\author[0000-0003-4034-0416]{Travis S. Metcalfe}
\affiliation{Center for Solar-Stellar Connections, White Dwarf Research Corporation, 9020 Brumm Trail, Golden, CO 80403, USA}
\email{travis@wdrc.org}

\begin{abstract}
For nearly a decade, observations have shown that many older Sun-like stars spin faster than predicted, a phenomenon known as weakened magnetic braking (WMB). The leading hypothesis for WMB is a weakening of the large-scale dipole field, which leads to a less efficient angular momentum loss. To test this hypothesis on a star known to be in the WMB regime, we present the first Zeeman Doppler Imaging (ZDI) map of the Sun-like star $\tau$\,Ceti, reconstructed using spectropolarimetric data from the Canada-France-Hawai'i Telescope (CFHT). Our ZDI analysis reveals a remarkably simple, stable and weak ($\langle B\rangle =0.17\,\mathrm{G}$) magnetic field, characterized by a predominantly dipolar ($\sim92$\% magnetic energy contained in $l=1$ modes), and highly axisymmetric ($\sim88$\% magnetic energy contained in $m<l/2$ modes) morphology. We infer a dipole field strength of $B_{\mathrm{dip}}=0.31\,$G, nearly an order of magnitude weaker than standard braking model predictions, providing direct confirmation of the weakened large-scale dipole predicted by the WMB hypothesis. This work establishes a new benchmark for ZDI, demonstrating that even extremely quiet stars in the WMB regime are accessible to this
technique.

\end{abstract}


\section{Introduction} 
Cool stars like the Sun possess convection zones that, when coupled with stellar rotation, generate substantial surface magnetic fields. These fields are fundamental to stellar evolution, influencing the star itself and its surrounding environment. Stellar magnetic fields are not static; they are complex structures that evolve across a wide range of spatial and temporal scales.

Sun-like stars are born with relatively rapid rotation that slows over time as magnetized stellar winds carry away angular momentum—a process known as magnetic braking \citep{parker1955,weber1967}. Over the past decade, the \textit{Kepler} mission has provided an unprecedented quantity of rotation period measurements, revealing discrepancies with the standard model of stellar spin-down. Many Sun-like stars remain unexpectedly rapid rotators at old ages \citep{angus2015}, suggesting that the efficiency of magnetic braking decreases dramatically after they pass the middle of their main-sequence lifetimes and reach a critical Rossby number \citep{ vansaders2016}, defined as the ratio between the rotation period, $\mathrm{P_{rot}}$, and the convective overturn timescale, $\tau_{cz}$ (Ro$=\mathrm{P_{rot}/\tau_{cz}}$; $\mathrm{Ro_{crit}\sim Ro_{\odot}}$). The coincident disappearance of Sun-like activity cycles at a similar age and Ro strongly suggests a magnetic origin for this ``weakened magnetic braking" \citep[WMB;][]{vansaders2016,metcalfe2017}. 

The leading hypothesis for WMB is a fundamental shift in the star's magnetic field strength and morphology. Because the large-scale dipole component of the magnetic field provides the most efficient lever arm for angular momentum loss \citep{reville2015,see2019}, a weakening of this component accompanied by a shift in magnetic energy toward smaller, more complex spatial scales would drastically reduce the braking torque \citep{garraffo2016,garraffo2018}. This hypothesis has been supported by spectropolarimetric snapshots showing that less active, older stars beyond $\mathrm{Ro_{crit}}$ exhibit much weaker large-scale dipole fields than their more active counterparts \citep{metcalfe2021,metcalfe2022,metcalfe2023,metcalfe2024a,metcalfe2025a}.
\begin{table*}[ht]
    \centering
    \caption{Log of the ESPaDOnS spectropolarimetric observations of $\tau$\,Ceti. The columns provide: (1) the UT date of the observation; (2) the rotational phase, calculated relative to the first epoch and using a rotation period of P$_{\mathrm{rot}}=34$ days \citep{baliunas1996}; (3) the number of exposures per night; (4) the mean S/N per pixel at 550 nm in the unnormalized Stokes $I$ spectrum; (5) the polarimetric precision in the coadded Stokes $V$ LSD profile, computed as the standard deviation of the coadded null LSD profile in the region of the spectral line; (6) the longitudinal magnetic field strength, $\langle B_{\rm z}\rangle$, computed from the coadded LSD Stokes $V$ profile; and (7) the False Alarm Probability (FAP) of the detection in the coadded Stokes $V$ LSD profile.}
    \begin{tabular}{ccccccc}
    \hline
    \hline
        UT Date & Phase & Num& S/N & $\sigma_V \,[\times10^{-6}]$ & $B_{\rm z}\pm\sigma_{B_{\rm z}}$ [G] & FAP \\
    \hline
    \hline
         2023-10-21& 0.00 &14 & 1014 & $6.26$& $-0.19\pm0.13$& $3.0\times10^{-3}$\\
         2023-12-02& 1.23 &15 & 990 & $5.38$& $-0.15\pm0.10$ & $5.5\times10^{-5}$\\
         2023-12-04& 1.29 &13& 1067 & $3.54$ & $-0.28\pm0.10$& $2.0\times10^{-5}$\\
         2023-12-30& 2.05 &14 & 1058 & $4.26$& $-0.28\pm0.09$& $1.3\times10^{-12}$\\
         2024-01-06& 2.26 & 15 &  1079 & $5.69$ & $-0.34\pm0.09$& $5.2\times10^{-7}$\\
         2024-01-16& 2.55 & 12 & 1133 & $5.48$& $-0.27\pm0.14$& $2.7\times10^{-3}$\\
    \hline
    \end{tabular}
    \label{tab1}
\end{table*}
An ideal laboratory to investigate this transition is $\tau$\,Ceti, a nearby \citep[$3.652 \pm 0.002$ pc;][]{gaia2022}, G8V \citep{keenan1989} and old \citep[4.4-12.4 Gyr;][]{lachaume1999,pijpers2003,difolco2004,mamajek2008,baum2022} star. A recent spectropolarimetric snapshot \citep{metcalfe2023} from the Large Binocular Telescope (LBT) revealed that its magnetic field is dominated by a weak, axisymmetric dipole component (mean longitudinal magnetic field $\langle B_{\rm z}\rangle=-0.37\pm0.08$ G, dipolar field strength $B_{\rm dip}=-0.77\pm0.31$ G), and an estimate of its wind-braking torque places it firmly in the WMB regime, making it one of the oldest stars \citep[$9.0\pm1.0\,$Gyr;][]{tang2011} yet identified in this state. The star's significance is further enhanced by its known planetary system, which includes at least four planets, two of which reside near the habitable zone \citep{feng2017}. Understanding the host star's magnetic environment is therefore integral for assessing the conditions for life on these worlds.

A single snapshot, however, provides only a limited view of a star's global magnetic field. To fully characterize its magnetic morphology and confirm its role in the WMB paradigm, observations at multiple rotational phases are required. In this Letter, we present new observations that enable the first Zeeman Doppler Imaging (ZDI) map of the global magnetic field of $\tau$\,Ceti.

\section{ESPaDOnS Observations}
Spectropolarimetric observations of $\tau$\,Ceti were conducted between 2023 October 21 and 2024 January 16 using ESPaDOnS at the 3.6m CFHT on Maunakea. In total, we acquired data from six individual epochs, with five of these observations spanning a 45-day period to facilitate rotational phase mapping.

ESPaDOnS is a high-resolution spectropolarimeter covering a wavelength range of 370 to 1050 nm with a spectral resolving power of R = 65,000 \citep{donati2003}. Each observation consisted of a standard polarimetric sequence of four consecutive sub-exposures, with each sub-exposure using a different orientation of the instrument's retarder waveplate. Operating in circular polarization mode, these sequences yield both the total intensity spectrum (Stokes $I$), derived by summing the four sub-exposures, and the circularly polarized spectrum (Stokes $V$), which is calculated from the sub-exposures with orthogonal polarization states. This mode is particularly well-suited for studying low-activity, cool dwarf stars like $\tau$\,Ceti, as Zeeman signatures present a larger amplitude in circular polarization compared to linear polarization \citep{landi1992}.

The raw data were processed using the \texttt{LIBRE-ESPRIT} reduction pipeline \citep{donati1997}. The reduced Stokes $I$ spectra of $\tau\,$Ceti show a typical signal-to-noise ratio (S/N) per pixel at 550 nm of 1060. The observation details are given in Table \ref{tab1}.
\section{Multi-line Polarization Analysis}
To enhance the S/N of the faint polarization signatures expected from $\tau$\,Ceti, we employed a multi-line analysis technique. This method leverages the redundant information contained within the thousands of metallic lines covered by modern echelle spectropolarimeters like ESPaDOnS. Specifically, we utilized Least-Squares Deconvolution \citep[LSD;][]{donati1997,kochukhov2010} to combine these lines into a single, high-S/N mean profile for both intensity (Stokes $I$) and circular polarization (Stokes $V$).

The LSD technique models an observed spectrum as a convolution of a predetermined line mask—representing the positions and relative strengths of selected atomic lines—with a single, common profile shape. Through a weighted linear least-squares process, this model is fit to the data to derive the high-precision LSD profile, which serves as a proxy for the average line profile across the stellar disk. Since LSD assumes that all lines have similar profile shapes, lines that are significantly broader than the average (e.g., hydrogen Balmer lines, Na D) must be excluded.

For the purpose of LSD profile calculations, we began the analysis by constructing a line mask tailored to $\tau$\,Ceti. We queried the Vienna Atomic Line Database \citep[VALD;][]{ryabchikova2015} to generate a list of 23,956 atomic transitions between 360–1010 nm. This list was then filtered to include only lines with a central depth greater than 1\% of the continuum, calculated for a MARCS model atmosphere \citep{gustafsson2008} with stellar parameters matching those of $\tau$\,Ceti ($T_{\mathrm{eff}}=5250$ K, log $g=4.5$, [M/H]$=-0.44$, $v_{\mathrm{micro}}=1.0$ km$\,\mathrm{s}^{-1}$), as reported in \cite{metcalfe2023}. After excluding regions affected by telluric contamination and the hydrogen Balmer series, our final line mask contained 5,219 metallic lines.

\begin{figure}
    \centering
    \includegraphics[width=\linewidth]{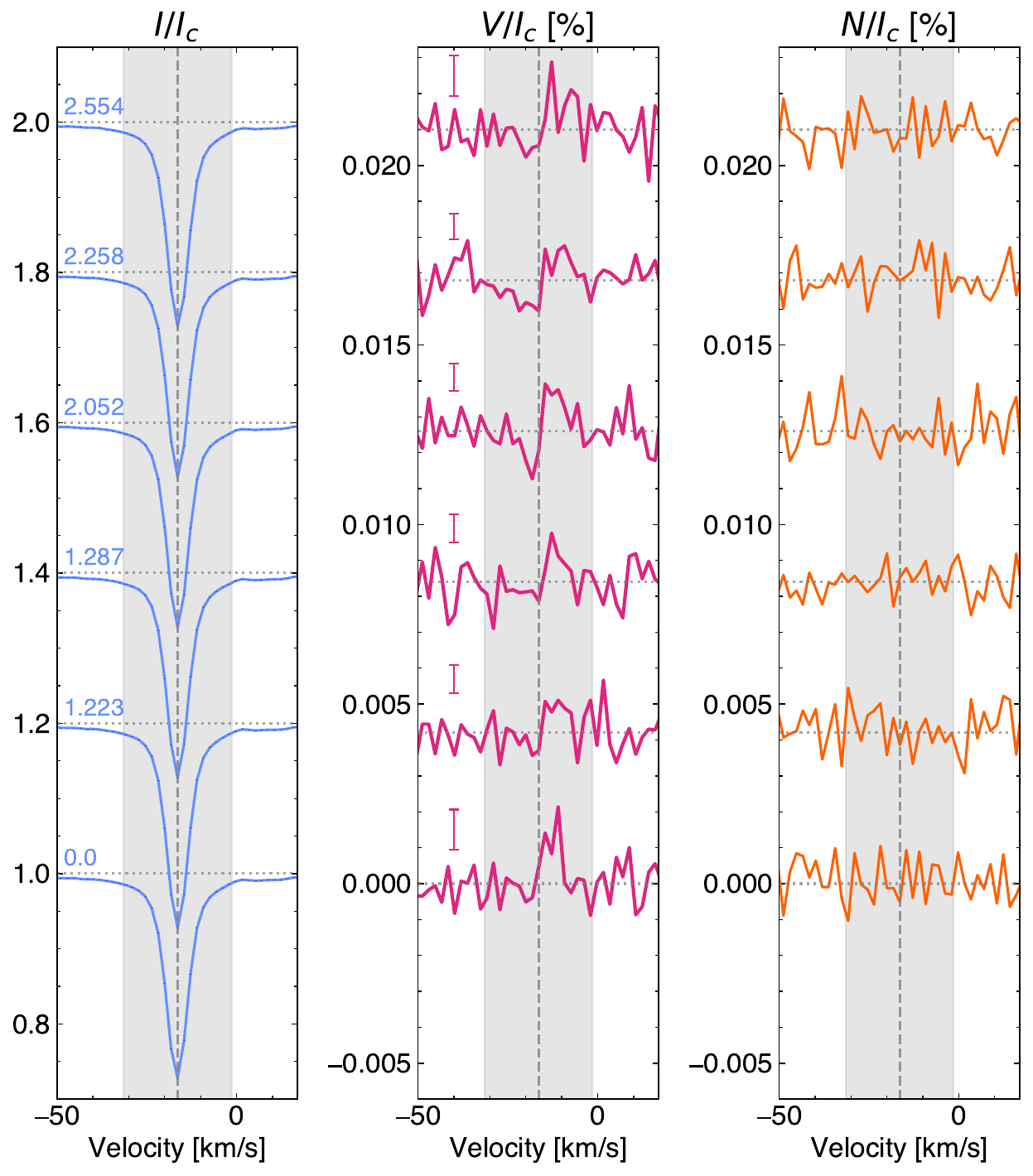}
    \caption{LSD profiles of $\tau$\,Ceti derived from ESPaDOnS data. The Stokes $I$ (left), $V$ (middle) and null profile (right) are offset vertically for display purposes. The different rotational phases are shown in the Stokes $I$ panel. Percentage mean errors on the computed LSD  Stokes $V$ profile are indicated with an error bar in the middle panel.}
    \label{fig1}
\end{figure}

We derived LSD profiles using \texttt{SpecpolFlow}\footnote{\url{https://folsomcp.github.io/specpolFlow/}} \citep{folsom2025}, a new, open-source pipeline. The process began with interactive continuum normalization of each ESPaDOnS spectrum following the procedure outlined in \citet{folsom2008} and \citet{folsom2013}. Based on the custom line mask, we used \texttt{SpecpolFlow} to compute LSD profiles for the Stokes $I$, $V$, and diagnostic null ($N$) spectra for each observation, following the methods of \cite{donati1997} and \cite{kochukhov2010}. The $N$ profiles were calculated by combining the four sub-exposures in such a way that polarization cancels out, allowing us to verify that no instrumental artifacts are present in the data. The LSD profiles were computed over a velocity range of $\pm200\, \mathrm{km\,s}^{-1}$ with a 1.8 $\mathrm{km\,s}^{-1}$ velocity step, matching the instrument's mean wavelength sampling.
Normalization weights for the LSD calculation were set to a central depth $d_0 = 0.4$, effective Landé factor $z_0=1.2$, and wavelength $\lambda_0=500$ nm. LSD profiles obtained from spectra taken during the same night were co-added using a $1/\sigma^2$ weighted average to produce a final, higher S/N LSD profile, as shown in Fig. \ref{fig1}.
From the final Stokes $V$ and $I$ profiles, we derived the mean longitudinal magnetic field, $\langle B_{\rm z}\rangle$, using the first-moment technique \citep{rees1979,donati1997,kochukhov2010}. To quantify the statistical significance of any magnetic detection, we used \texttt{SpecpolFlow} to calculate the FAP—the probability that the observed Stokes $V$ signal is consistent with noise. Following the convention adopted in \cite{donati1997}, we define a ``definitive detection" as having a FAP$<10^{-5}$ and a ``marginal detection" as having a FAP$<10^{-3}$.

We obtain two definite detections (FAP $<10^{-5}$) of the circular polarization signature in Stokes $V$ LSD profiles (UT Date 2023-12-30 and 2024-01-06), two marginal detections (FAP $<10^{-3}$; UT Date 2023-12-02 and 2023-12-04) and two non-detections (FAP $>10^{-3}$; UT Date 2023-10-21 and 2023-01-16). The overall polarity of the Stokes $V$ profiles is negative and we infer a weighted mean longitudinal magnetic field strength $\langle B_{\rm z} \rangle = -0.26\pm0.04$ G. A detailed list of $\langle B_{\rm z} \rangle$ and FAP for each coadded LSD profile is reported in Table \ref{tab1}. 
\section{Zeeman Doppler Imaging}
To reconstruct the surface magnetic field geometry of $\tau$\,Ceti, we used the Zeeman Doppler Imaging (ZDI) code \texttt{InversLSD} \citep{kochukhov2014}, which has been extensively validated and used in numerous studies of stellar magnetic field morphology \citep[e.g.,][]{rosen2015,kochukhov2017,oksala2018,kochukhov2019,kochukhov2022,kochukhov2025}.

ZDI is an inversion technique that iteratively fits a model of a star's surface to a time series of observed Stokes profiles. The process begins by dividing the stellar surface into a grid of independent zones—1,876 in this work following the optimal surface sampling scheme discussed in \citet{piskunov2002}. We then calculated synthetic local Stokes $I$ and $V$ profiles for each zone, which we subsequently integrated over the visible stellar disk at each observed rotational phase. We compared these disk-integrated synthetic profiles to our observed LSD profiles, and the surface map is updated in successive iterations until an optimal fit is achieved.

As ZDI is an inherently ill-posed problem, robust regularization is needed to converge on a unique and physically plausible solution. The magnetic field was modeled using a spherical harmonic expansion, which decomposes the field into its radial poloidal (specified by the harmonic coefficient $\alpha_{lm}$, with angular degree, $l$, and azimuthal order of each mode, $m$), horizontal poloidal ($\beta_{lm}$) and horizontal toroidal ($\gamma_{lm}$) components \citep{kochukhov2014}. We regularized the magnetic inversion with a harmonic penalty function, $\Lambda\Sigma_{lm}l^2(\alpha_{lm}^2+\beta_{lm}^2+\gamma_{lm}^2)$, which favors simpler, large-scale field geometries by suppressing power in high-order modes. The contribution of this regularization function is controlled by a regularization parameter, $\Lambda$. We determined the optimal value of $\Lambda$ iteratively. We began with a high value of regularization and progressively reduced it until the model's fit to the observations no longer showed significant improvement \citep{kochukhov2017a}.

We parametrized the global magnetic field geometry of $\tau$\,Ceti using a spherical harmonic expansion that included both poloidal and toroidal terms. The spherical harmonic expansion was truncated at a maximum angular degree of $l_{\mathrm{max}}=5$, since the observer is blind to magnetic field structures at smaller scales for slow-rotators ($v\,\mathrm{sin}\,i<5\,\mathrm{km\,s^{-1}}$) like the Sun and $\tau$\,Ceti \citep{lehman2019}. The local Stokes profiles used in the inversion were pre-calculated using the Unno-Rachkovsky analytical solution of the polarized radiative transfer equation \citep{landi2004}, for magnetic field strengths from 0 to 50 G in 2 G increments, across 15 limb angles and 15 field vector inclinations. For the ZDI modeling, we adopted several stellar parameters from the literature, including a projected rotational velocity \citep[$v\,\mathrm{sin}\,i=0.4\,\mathrm{km\,s^{-1}}$;][]{saar1997}, a radial velocity \citep[$v_{\mathrm{rad}}=-16.4\,\mathrm{km\,s^{-1}}$;][]{hourihane2023} and an inclination angle \citep[$i=20^{\circ}$, as calculated by ][from the values of $v\,\mathrm{sin}\,i$, stellar radius $R=0.816\,R_{\odot}$ \citep{vonbraun2017} and P$_{\mathrm{rot}}=34$ days \citep{baliunas1996}]{metcalfe2023}. 
\begin{figure*}[ht!]
    \centering
    \includegraphics[width=\linewidth]{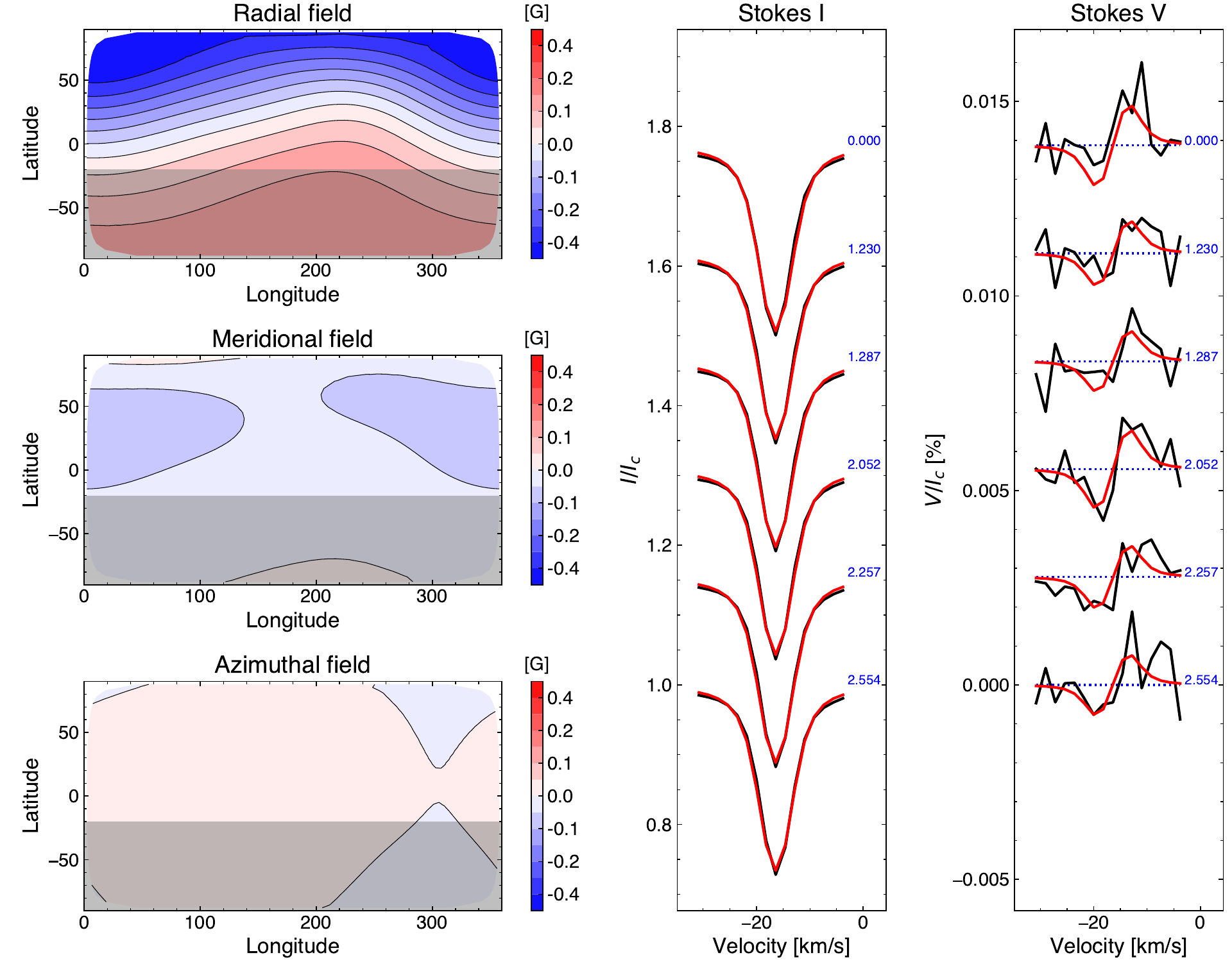}
    \caption{Result of the ZDI analysis of $\tau$\,Ceti. The three rectangular maps represent the surface distributions of the radial, meridional, and azimuthal magnetic field components. The color bar indicates the polarity and strength (in G) of the magnetic field. The grey shaded area at southern latitudes represents the portion of the stellar surface that remains hidden from view due to the star's low inclination. The two vertical panels show a comparison between the observed (black) and model (red) Stokes $I$ and $V$. The spectra are offset vertically, with the rotational phase indicated next to each profile.}
    \label{fig2}
\end{figure*}
The ZDI map of $\tau$\,Ceti is presented in Fig.~\ref{fig2}. The surface-averaged field modulus $\langle B\rangle=0.17\,$~G, with the poloidal component accounting for 99\% of the total magnetic energy. The dipolar mode accounts for $\sim92\%$ of the poloidal component and both the quadrupolar and octupolar modes represent a very small fraction, namely 6.4\% and 0.6\%. The field is predominantly axisymmetric ($\sim88\%$), with the dipolar mode contributing $\sim 82\%$ to the global axisymmetry and the quadrupolar and octupolar components contributing 4.8\% and 0.6\%, respectively. Using the $\alpha_{1,m}$ harmonic coefficients, we inferred a dipolar field strength of $B_{\rm dip}=0.31$~G and an obliquity of $\beta=160.13\degr$. The standard deviation between the observed and model Stokes $I$ and $V$ profiles is $\sigma_{I}=8.4\times10^{-3}$ and $\sigma_{V}=8.0\times10^{-6}$. Note that the derived magnetic field strength does not depend strongly on the inclination (e.g., adopting the inclination of $\tau\,$Ceti's debris disk of 35$^{\circ}$ from \cite{bisht2024}, we find $\langle B\rangle=0.18\,$G and $B_{\mathrm{dip}}=0.33\,$G).
\section{Discussion \& Conclusions}
In this Letter, we present the first large-scale magnetic field geometry map of the old Sun-like star $\tau$\,Ceti, reconstructed using a series of high-resolution spectropolarimetric observations from ESPaDOnS. Our Zeeman Doppler Imaging analysis reveals a magnetic field that is remarkably simple, characterized by a weak ($\langle B\rangle = 0.17$ G), predominantly dipolar ($\sim92$\%), and highly axisymmetric ($>82$\%) morphology. 

We find that $\langle B_z \rangle$, the longitudinal field averaged over the \textit{visible disk}, is larger than $\langle B\rangle$, the field modulus averaged over the \textit{entire stellar surface}. This seemingly unexpected result (for a pure dipole) stems from a combination of two factors. First, the star's nearly pole-on inclination and concentration of negative field in the visible pole maximize $\langle B_z \rangle$. Second, our ZDI inversion is not restricted to fitting a pure dipole ($\alpha_{1,m}=\beta_{1,m}$,  $\gamma_{l,m}=0$). Instead, we use the field standard-approach of fitting a general harmonic field morphology, which implies fitting and regularizing the spherical harmonic coefficients for the radial ($\alpha_{l,m}$) and horizontal ($\beta_{l,m},\gamma_{l,m}$) field components independently. In this case, the resulting small deviations from a pure dipole -- both $l \ge 2$ harmonic terms and $\beta\neq\alpha$, $\gamma\neq0$ degrees of freedom -- conspire to decrease the inferred $\langle B \rangle$ value relative to the observed $\langle B_z \rangle$. While some of these non-dipolar structural features of the field geometry may not be reliable, the usage of the general harmonic parameterization is required for comparison of our results to other ZDI studies.

Our findings robustly confirm and significantly extend the results from a previous spectropolarimetric snapshot of $\tau$\,Ceti obtained with the LBT \citep{metcalfe2023}. While that single observation yielded only a marginal detection, our multi-epoch dataset provides definitive detections of the star's circularly polarized signature at multiple rotational phases. These two independent datasets are in excellent quantitive agreement: the Stokes $V$ amplitude in our LSD profiles is 15–20 ppm (Fig.~\ref{fig1}), and the mean longitudinal magnetic field is $\langle B_{\rm z} \rangle=-0.26\pm0.04$ G, both in close agreement with the LBT snapshot values ($\sim 20$ ppm and $\langle B_{\rm z} \rangle=-0.37 \pm 0.08$ G, respectively). Furthermore, our ZDI-inferred dipolar field strength of 0.31 G is consistent with the $B_\mathrm{dip}=0.77\pm0.31\,$G value inferred by \cite{metcalfe2023} using a simpler forward-modeling approach assuming an axisymmetric dipole morphology. The remarkable consistency of the field's strength and dominant negative polarity over a 16-month baseline points to a highly stable magnetic configuration. This stability is expected, given $\tau$\,Ceti's lack of a known activity cycle and its nearly pole-on viewing orientation ($i\approx20^{\circ}$), which provides a persistent view of the star's polar regions. 

\begin{figure*}[ht!]
    \centering
    \includegraphics[width=\linewidth]{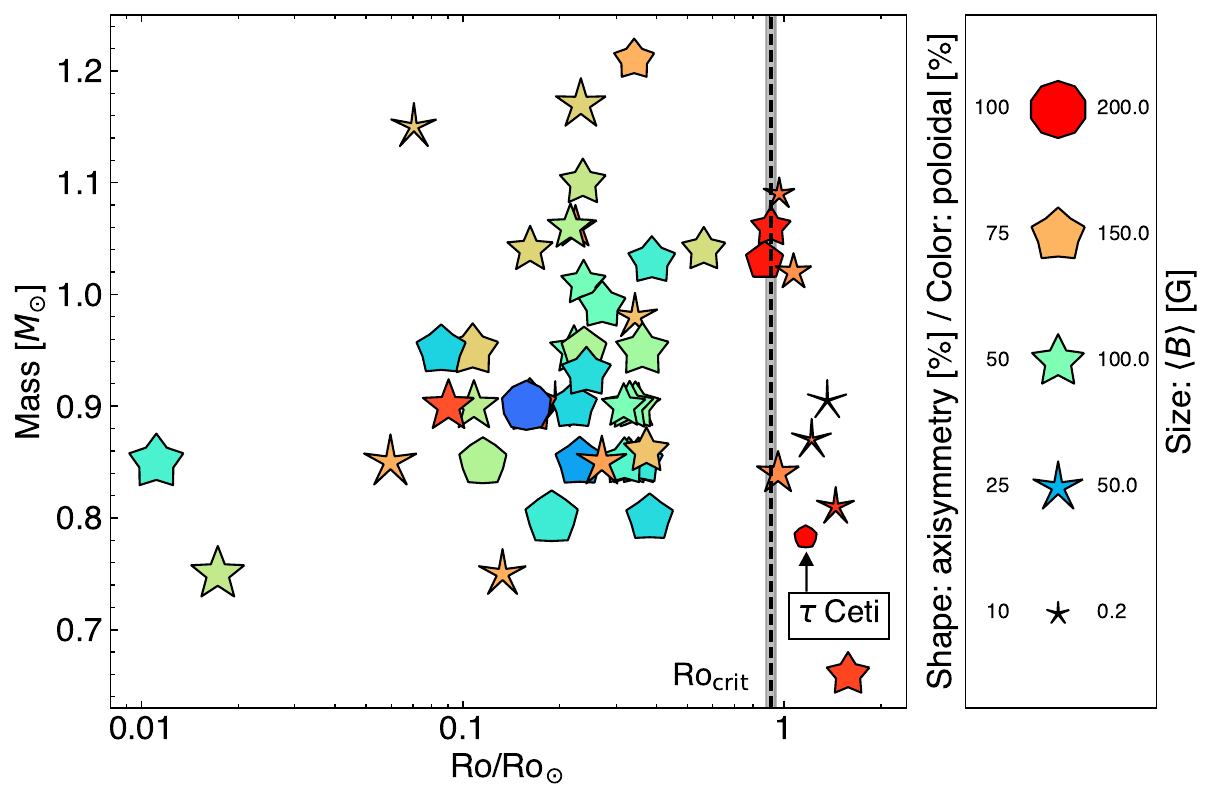}
    \caption{Properties of the global magnetic morphologies for cool, main sequence stars obtained via ZDI. All values of Rossby numbers are calculated from the Gaia $G_{\mathrm{BP}}-G_{\mathrm{RP}}$ color using the asteroseismic calibration of \cite{corsaro2021}, normalized by the solar value on this scale (Ro$_{\odot}=0.496$). The location of $\tau$\,Ceti is highlighted with an arrow and labeled. The critical Rossby number, Ro$_{\mathrm{crit}}=0.91\pm0.03$ \citep{saunders2024}, is shown as a vertical, dashed black line, with a shaded gray band representing its uncertainty. Data for $\tau$\,Ceti are from this work. All other data are from: \cite{donati2008,morgenthaler2012,jeffers2014,waite2015,borosaikia2015,borosaikia2016,folsom2016,donascimento2016,borosaikia2018,folsom2018a,folsom2018b,see2019b,folsom2020,brown2021,willamo2022,bellotti2024,metcalfe2024a,bellotti2025,metcalfe2025a}.}
    \label{fig3}
\end{figure*}

Placing these results in the broader context of stellar magnetic and rotational evolution, the properties of $\tau$\,Ceti are consistent with the trends observed for a limited but growing sample of Sun-like stars with ZDI maps. As shown in Fig.~\ref{fig3}, as stars evolve past the critical Rossby number and enter the WMB regime, their magnetic energy appears to become increasingly concentrated in the poloidal component, and their overall field strength decreases, just as revealed by the ZDI map of $\tau$\,Ceti. We find no clear trend between stellar mass, Ro and level of axisymmetry. 

Our ZDI map reveals a magnetic field strength for $\tau$\,Ceti that is significantly weaker than predicted by \textit{standard} braking models. Assuming that the magnetic field strength scales with photospheric pressure $P_{\mathrm{phot}}$ and Rossby number as $B/B_{\odot}=(P_\mathrm{phot}/P_\mathrm{{phot,\odot}})^{1/2}\,\mathrm{(Ro/Ro_{\odot})}^{-1}$ \citep{vansaders2013}, and using the values of $P_{\mathrm{phot}}$ and Ro computed from a standard braking model \citep{vansaders2013}, Ro/Ro$_{\odot}$ = 1.05 and $B/B_{\odot}$ = 1.21 for $\tau$\,Ceti \citep{metcalfe2025b}. Adopting the strength of the solar dipole field ($B_{\mathrm{dip},\odot}=1.54\pm0.66$ G) from \cite{Finley2018}, this implies an expected dipole field strength of $\sim 2\,$G for $\tau$\,Ceti. This prediction is nearly an order of magnitude larger than the $B_{\mathrm{dip}} = 0.31\,$G we infer from our ZDI map. This discrepancy reinforces the central premise of the WMB hypothesis: that the large-scale dipole field of old Sun-like stars is weaker than standard models assume.

During WMB, \cite{metcalfe2016} has suggested that magnetic energy is preferentially concentrated into smaller spatial scales, which are less efficient at braking the star, thus allowing old Sun-like stars to retain faster rotation than expected from standard spin-down models \citep{vansaders2016}. We do not detect high-order components in the magnetic field of $\tau$\,Ceti. However, we argue that the lack of such detections is not at odds with the hypothesis of \cite{metcalfe2016}, but rather exposes the inherent biases of ZDI when observing a target with the physical properties of $\tau$\,Ceti. By applying ZDI to magnetic field simulations, \cite{lehman2019} demonstrated that ZDI is able to recover the main magnetic structures of the large-scale field (e.g. dipole, quadrupole and octupole) for Sun-like stars, especially after averaging over several maps. However, they also showed that ZDI systematically struggles to recover high-order components of the magnetic fields of low-inclination, slowly-rotating and weakly active Sun-like stars \citep{lehman2019}. In particular, the combination of a low inclination angle (pole-on view) and slow rotation leads to low values of  $v\,\mathrm{sin}\,i$. The lower $v\,\mathrm{sin}\,i$ yields a smaller range of Doppler-shifted contributions across the stellar surface, which decreases the amount of spatial information encoded in the corresponding spectral lines. At $v\,\mathrm{sin}\,i=0.4\,\mathrm{km\,s^{-1}}$ of $\tau$\,Ceti, we are limited by the thermal width of the line profile as well as the spectral resolution of the instrument. The cumulative result, as found by \cite{lehman2019}, is that ZDI can underestimate the true magnetic energy by $\sim$ one order of magnitude, with the energy tending to decrease with increasing $l$-mode (within the allowed range of $l$ modes, i.e $l_{\mathrm{max}}=5$ in this study). Therefore, the absence of smaller-scale fields predicted for stars in the WMB regime is likely a consequence of observational limitations, rather than a true physical absence of such fields on $\tau\,$Ceti.

In conclusion, this work establishes a new benchmark for ZDI, demonstrating that even extremely quiet stars in the WMB regime, like $\tau$\,Ceti, are accessible to this technique. We presented the first ZDI map of $\tau$\,Ceti, constraining its surface-averaged field strength to only $\langle B\rangle=0.17\,$G--the weakest yet inferred with ZDI--and revealing a stable, and predominantly axisymmetric, dipolar field. Our results confirm that stars in the WMB regime possess extremely weak large-scale fields, but they simultaneously underscore the significant challenge of testing the full hypothesis regarding the redistribution of magnetic energy into higher-order modes \citep{vansaders2016}. A systematic ZDI survey of Sun-like stars across a range of Rossby numbers, masses, and compositions will help understand the behavior of stars before, during, and after this magnetic transition.

\section*{Acknowledgments}

\noindent We thank the anonymous referee for their careful review and valuable feedback, which greatly improved the quality of this Letter. F.C. acknowledges Dhvanil Desai for helpful discussions. F.C and J.v.S. acknowledge support from NSF grant AST-2205888. O.K. acknowledges funding by the Swedish Research Council (grant agreement 2023-03667) and the Swedish National Space Agency.
T.S.M.\ acknowledges support from NSF grant AST-2205919.
This work has made use of the VALD database, operated at Uppsala University and the University of Montpellier. We thank Dr. T. Ryabchikova for her invaluable work with compiling and assessing atomic data for the VALD database. Based on observations obtained at the Canada-France-Hawaii Telescope (CFHT) which is operated by the National Research Council (NRC) of Canada, the Institut National des Sciences de l'Univers of the Centre National de la Recherche Scientifique (CNRS) of France, and the University of Hawaii. The observations at the CFHT were performed with care and respect from the summit of Maunakea which is a significant cultural and historic site.

\facilities{CFHT (ESPaDOnS)}
\software{\texttt{InversLSD} \citep{kochukhov2014} \texttt{SpecpolFlow} \citep{folsom2025}; 
          }

\bibliography{main}{}
\bibliographystyle{aasjournal-compact}

\end{document}